\documentclass{article}

\usepackage{spconf,amsmath,graphicx, enumitem,hyperref,booktabs,tabto,xcolor,multirow,bbding,breakurl}

\title{towards a unified Conformer structure: from ASR to ASV task}
\name{Dexin Liao$^1$, Tao Jiang$^{\dag3\thanks{\dag\: Co-first author \quad *\: Corresponding author}}$, Feng Wang$^1$, Lin Li$^{*2}$, Qingyang Hong$^{*1}$}
\address{$^1$School of Informatics, Xiamen University, China\\
	$^2$School of Electronic Science and Engineering, Xiamen University, China \\
         $^3$Xiamen Talentedsoft Co., Ltd., China}
\email{\{lilin,qyhong\}@xmu.edu.cn}
\begin{document}
\ninept
\maketitle
\begin{abstract}
Transformer has achieved extraordinary performance in Natural Language Processing and Computer Vision tasks thanks to its powerful self-attention mechanism, and its variant Conformer has become a state-of-the-art architecture in the field of Automatic Speech Recognition (ASR). However, the main-stream architecture for Automatic Speaker Verification (ASV) is convolutional Neural Networks, and there is still much room for research on the Conformer based ASV. In this paper, firstly, we modify the Conformer architecture from ASR to ASV with very minor changes. Length-Scaled Attention (LSA) method and Sharpness-Aware Minimization (SAM) are adopted to improve model generalization. Experiments conducted on VoxCeleb and CN-Celeb show that our Conformer based ASV achieves competitive performance compared with the popular ECAPA-TDNN. Secondly, inspired by the transfer learning strategy, ASV Conformer is natural to be initialized from the pretrained ASR model. Via parameter transferring, self-attention mechanism could better focus on the relationship between sequence features, brings about 11\% relative improvement in EER on test set of VoxCeleb and CN-Celeb, which reveals the potential of Conformer to unify ASV and ASR task. Finally, we provide a runtime in ASV-Subtools to evaluate its inference speed in production scenario. Our code is released at \texttt{\url{https://github.com/Snowdar/asv-subtools/tree/master/doc/papers/conformer.md}}.
\end{abstract}
\section{Introduction}
\label{sec:intro}
Automatic Speaker Verification (ASV) is a task to verify the identity of the speaker by voice, which has been well-developed and widely applied in many real-world scenarios. Currently, x-vector proposed by Snyder et al. \cite{Snyder2018XVectorsRD} is the most popular framework for ASV systems. It includes two parts, where an embedding extractor maps utterances with variable duration to fixed-dimensional speaker representations, and then the similarity of the speaker representation can be calculated by back-end scoring method. Many prior works focused on DNN-based structure have improved the performance of ASV systems (e.g., ResNet, Res2Net, ECAPA-TDNN) \cite{2019BUT,2021ResNeXt,2020ECAPA,2021The}. Most of above networks are Convolutional Neural Networks (CNNs), which have the inherent ability of emphasizing the local information.

Recently, self-attention mechanisms that directly capture the global information have been explored, and it has helped Transformer \cite{NIPS2017_3f5ee243} achieve remarkable success in Natural Language Processing (NLP) and Computer Vision (CV) areas \cite{devlin2018bert, dosovitskiy2020vit}. However, unlike CNNs, Transformer lacks some of the inductive biases, such as translation equivariance and locality, which degrades performance when trained on insufficient data. It is difficult to achieve competitive results by directly applying Transformer to ASV tasks \cite{t-v, local_tr}. Conformer \cite{conf} is a hybrid architecture which combines self-attention with convolutions, i.e., self-attention learns the global interaction while convolutions capture the local information. It has become a state-of-art model in Automatic Speech Recognition (ASR). MFA-Conformer \cite{mfa} utilizes the Multi-scale Feature Aggregation method in ECAPA-TDNN, successfully introduces Conformer into ASV for the first time. However, the uniformity between ASV an ASR deserves further attention. The same Conformer establishes connections between ASV and other tasks, which will not only facilitate better research on the link between ASV and ASR, but may also be a foundation of future multi-task learning or multimodal machine learning. Hence, in this paper, we mainly concentrate on ASV Conformer which matches ASR encoder. 

In addition, several studies have injected phonetic information into the DNN structure of the ASV extractor through multi-task learning \cite{mt-1, svkws,voi}, indicates that there exists some positive interdependence between the speaker identities and ASR tasks when sharing some of the low-level computation. Meanwhile, in the field of Language Identification (LID), providing informative speech representation by a pretrained ASR model in LID system, proved to be effective for the downstream LID task \cite{lid}. It is worth mentioned that \cite{ant} adopts transfer learning scheme, that is, pretrains a U2++ encoder-decoder \cite{u2} model and then further finetunes the encoder for the LID task, won the first place in the OLR 2021 \cite{olr}. It can be well explained by the fact that the ASR encoder already has a strong capability to discriminate languages, since the supervised training labels for ASR are language-related. Although the association with ASR information in ASV task is not as apparent as in LID, e.g., different speakers can say the same words. Their deeper dependencies could be digged by appropriate methods. Inspired by these works, we propose a parameter transferring strategy, which can make use of a typical ASR model to improve the performance of ASV system.

At last, for the purpose of bridging the gap between production and research, we provide a C++ based runtime tool to evaluate our models' inference speed in production environment. With Torch Just In Time (JIT) and LibTorch, models trained by Pytorch can be converted to TorchScrip, and then employed in C++ applications. Our main contributions in this paper are as follows.
\begin{itemize}[leftmargin=*]

\item 
We modify Conformers of different configures from ASR to ASV system. To improve model generalization ability, Length-Scaled Attention (LSA) method \cite{2022Overcoming} enables self-attention to better generalize to various length inputs, and Sharpness-Aware Minimization (SAM) \cite{foret2021sharpnessaware} prevents the loss from falling into the local minima during training. Our system yields competitive results in popular VoxCeleb \cite{Nagrani17,Chung18b} and CN-Celeb \cite{2019CN,LI202277}.
\item 
Through a parameter transferring strategy, we show that ASV Conformer could benefit from ASR information. Parts of the ASR encoder is selected to initialize ASV Conformer, then we retrain the model rather than finetune it. This method allows model to learn the deep relationship with ASR.
\item
We provide a runtime to conform the production value of our models, make it easier and more convenient to deploy ASV models to real applications. 

\end{itemize}

\section{Methods}
\label{sec:format}
\subsection{Model architecture}

\begin{figure}[htb]
  \centering
  \centerline{\includegraphics[width=8.5cm]{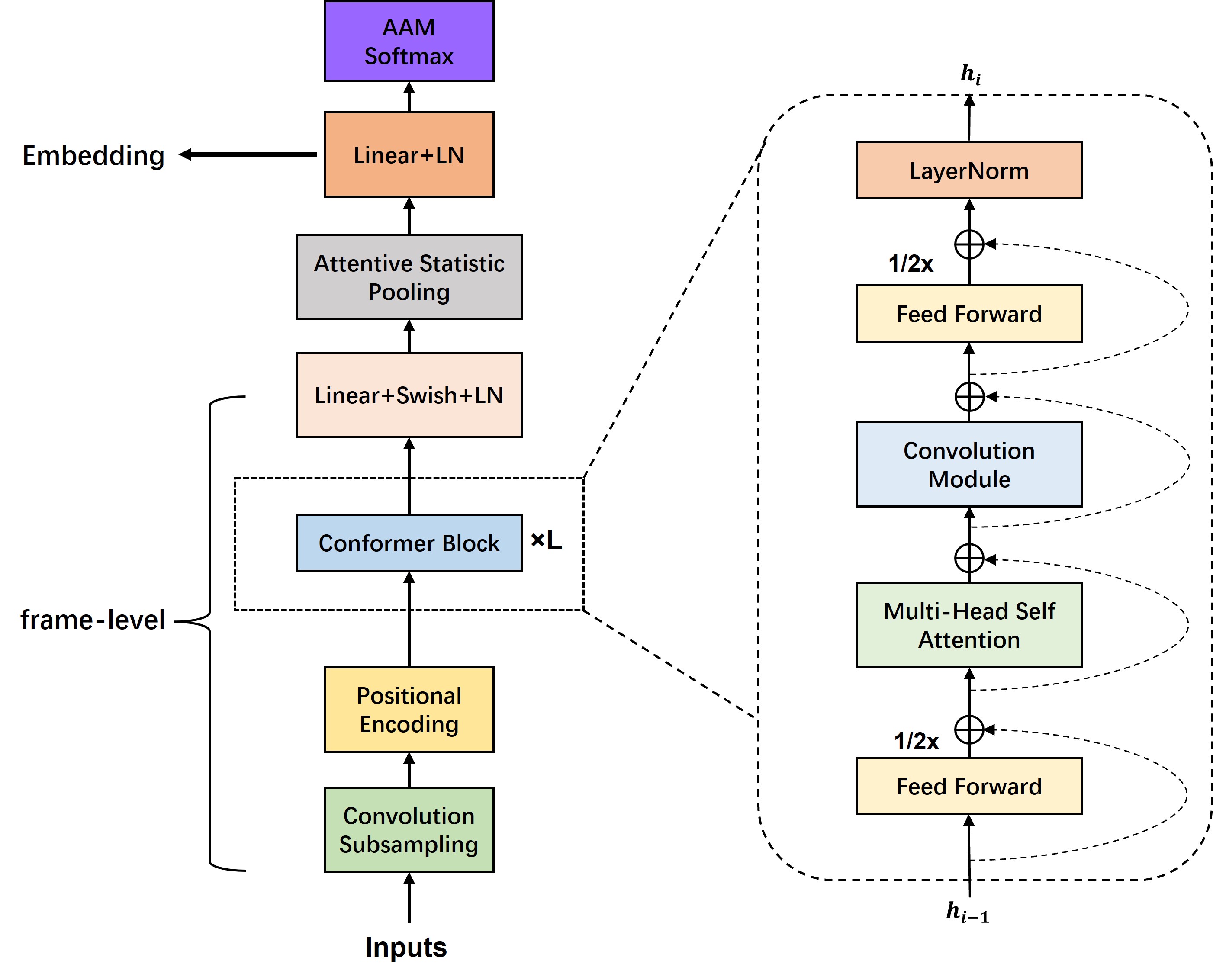}}
%
\caption{Schematic diagram of ASV Conformer.}
\label{fig:conv}
\vspace{-0.5em}
\end{figure}

The overview of ASV Conformer system is shown in Fig.~\ref{fig:conv}. Rotary Position Embedding \cite{su2022roformer}, which incorporates explicit relative position dependency in the form of absolution position embedding, encode the position information into the self-attention mechanism. On one hand, a stack of Conformer blocks model the frame-level speaker representation. On the other hand, an Attentive Statistics Pooling \cite{Okabe_2018} layer process all the information across the time dimension, resulting in a segment-level vector. After linear layer, the segment-level vector is further projected to fixed-dimensional x-vector. In training stage, the speaker embedding extractor is optimized by AAM-Softmax \cite{8953658} loss.

The Conformer block consists of two Macaron-like feed forward modules (FNN) with half residual connections, between them lies the Multi-head self-attention (MHSA) and Convolution module (Conv). MHSA means to model the global information, and Conv gives inductive bias to network.   

Without loss of generality, we assume MHSA has one attention head. An standard self attention can be described as mapping a query and a set of key-value pairs to an output, and the matrix of outputs is computed as the weighted sum over the value representation:
\begin{figure}[htbp]
  \centering
  \centerline{\includegraphics[width=8.5cm]{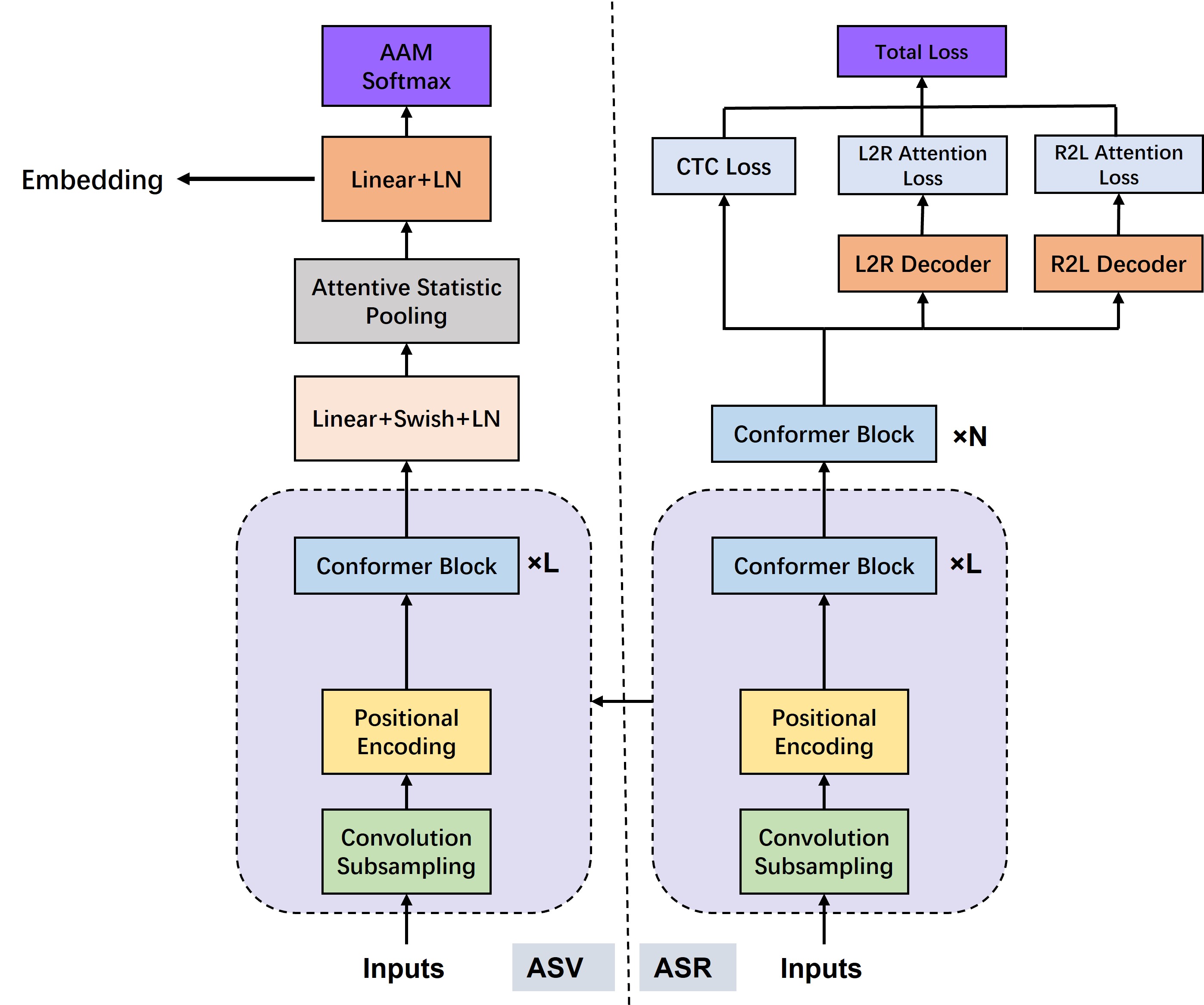}}
%
\caption{Schematic diagram of ASR transferring.}
\label{fig:trans}
\vspace{-0.5em}
\end{figure}
\begin{equation}
\operatorname{Att}(Q, K, V)=\operatorname{Softmax}\left(\frac{Q K^T}{\sqrt{d}}\right) V
\label{eq:att}
\end{equation}
where $d$ is the embedding dimension, and its detailed form is:
\begin{equation}
\boldsymbol{o}_i=\sum_{j=1}^n a_{i, j} \boldsymbol{v}_j, \quad a_{i, j}=\frac{exp \left({\frac{\boldsymbol{q}_i \cdot \boldsymbol{k}_j}{\sqrt{d}}}\right)}{\sum_{j=1}^n exp \left({\frac{\boldsymbol{q}_i \cdot \boldsymbol{k}_j}{\sqrt{d}}}\right) }
\vspace{-0.5em}
\end{equation}
where $n$ denotes the sequence length, $\boldsymbol{q}$, $\boldsymbol{k}$ and $\boldsymbol{v}$ correspond to query, key and value vector respectively, and $a_{i, j}$ means the attention score weight between time position $i$ and $j$. The softmax function normalizes the attention weight across all position to a probability distribution. However, there exits varying length inputs in inference stage, and this mismatching may hurt the original relationship of different position embeddings (e.g., a longer inputs diluting the attention weight, making the value in position $i$ is inclined to be easier to be influenced by unrelated position values). From the perspective of entropy \cite{kexuefm-9034}, equation~\ref{eq:att} is modified to a length-scaled version as equation~\ref{eq:att-e} to stabilize the uncertainty of attention distribution.

\begin{equation}
\operatorname{Att}(Q, K, V)=\operatorname{Softmax}\left(\frac{\log n }{s{\sqrt{d}}}{Q K^T}\right) V
\label{eq:att-e}
\end{equation}
where $n$ denotes the sequence length, $s$ is set to a learnable scalar as a temperature of softmax in different layers. In this way, the attention weights after softmax function are sharpened by $\log n$ in long sequences, while smoothed in short sequences.
\subsection{Sharpness-Aware Minimization training}

While Conformer's superior representational capacity enables ASV system to memorize the training set easily, it also leads to overfitting problems, especially when training data is insufficient. It has been studied \cite{sha, gen} that a model converging to sharp minimas of loss might results poorer generalization. Sharpness-Aware Minimization (SAM), which leverages the connection between geometry of the loss landscape and generalization, can seek a flatter minima, and is adopted to improve generalization ability of our models.

Intuitively, SAM aims to find the parameter $w$ whose entire neighbours have low training loss $L_{train}$, it can be defined as:
\begin{equation}
\min _w \max _{\|\epsilon\|_2 \leq \rho} L_{train }(w+\epsilon)
\end{equation}
where $\rho\geq 0$ denotes the radius of seeking region. A two-step approximation is applied to solve this minimax optimization:

\begin{equation}
\left\{\begin{array}{l}
\epsilon_t \approx \rho \nabla L_{train}\left(w_t\right)/\left\|\nabla L_{train}\left(w_t\right)\right\|_2 \\
\left.w_{t+1} \approx w_t-\alpha_ t\nabla L_{train}(w_t)\right|_{w_t+\epsilon_t}
\end{array}\right.
\end{equation}
where $\alpha_ t$ is the learning rate in training position $t$. The first step is an efficient approximation of $$\epsilon_t = \arg \max _{\|\epsilon\|_2 \leq \rho} L_{{train }}(w_t)+\epsilon_t^{T} \nabla_{w_t} L_{train}(w_t)$$
and in the second step, SAM updates weights based on the gradient in $w_t+\epsilon_t$.

\subsection{ASR transferring}
 
In order to integrate ASR information, we pretrain the U2++\cite{u2} ASR model first, and then optimize it to ASV task. However, unlike LID task \cite{ant}, ASR representation is not so related with speaker classification, so in the second stage we apply the same strategy as training the original ASV Conformer. In addition, it was previously found that ASR information is more helpful in shallow layers \cite{mt-1, svkws}, we infer that transferring parts of ASR encoder to ASV Conformer is compatible. Notably, the datasets used for ASR training are independent and the model configure is almost same as it in ASV, giving flexibility and convenient to the transfer scheme, i.e., a model for ASR task can be directly transferred to ASV. As depicted in Fig.~\ref{fig:trans}, the right part is U2++ architecture, which can be either an open source pretrained model or trained from scratch. Then the shared encoder is further to be trained to ASV task.

\section{Experimental setup}
\label{sec:pagestyle}

\subsection{Datasets}

We conducted our proposed experiments on CN-Celeb \cite{2019CN,LI202277} and VoxCeleb \cite{Nagrani17,Chung18b} respectively. The CN-Celeb corpus contains speech from Chinese celebrities. The entire dataset can be divided into CN-Celeb.T and CNC-Eval: the former involves 632,736 utterance from 2,793 speakers with total 1,285 hours, and is used for training; the latter is test set. The VoxCeleb is one of the most classic English dataset in the field of speaker verification. For the VoxCeleb, we only used the VoxCeleb2 which involves 1,092,009 utterance from 5994 speakers with total 2,300 hours as training set, and employed three available test trials: VoxCeleb1-O-clean, VoxCeleb1-E-clean and VoxCeleb-H-clean to verify our proposed method.

The pretrained ASR model involves three datasets, GigaSpeech \cite{GigaSpeech2021}, Multi-CN\cite{multicn} and Wenetspeech \cite{wenetspeech}. GigaSpeech is English speech recognition corpus with 10,000 hours, while Multi-CN and WenetSpeech are Chinese datasets with 2,825 and 10,005 hours respectively.
\subsection{Experimental Configuration}
We explored various Conformer configures, with a major emphasis on attention dimension, layer number and subsampling rate. The feed-forward dimension is set to 2048.

On-the-fly approach is adopted for data preprocessing. Data augmentation is applied to enrich the training data and no voice activity detection is performed. Data augmentation involves additive noise, reverberation and speed perturb. Additive noise from MUSAN \cite{musan} dataset is mixed with original signal, the room impulse responses from RIRs \cite{rir} dataset is used to inject reverberation via convolution operation. We extract 80-dimensional Mel-filterbanks as acoustic feature. Then ASV models are trained on chunks of 300 randomly selected from whole feature map. Cepstral mean-normalization (CMN) is applied before model training.

For ASV training, models are trained with AdamW optimizer with a total batch of 512 on 4 Nvidia V100 GPUs. The learning rate increases to a peak during the warmup stage and then decays to small as model converges. The total training process lasts about 30 epochs. For ASR pre-training, we follow the recipes in Wenet \cite{u2}.

More training details can refer to ASV-Subtools\cite{tong2021asv}.
\subsection{Model evaluation}
During the test, each utterance is chunked with about 300 frames. Embeddings extracted from chunks are averaged to the final x-vector. For back-end, we choose cosine similarity to score the extracted x-vectors. Evaluation performance is measured by Equal Error Rate(EER) and minimum normalized detection cost (minDCF).

From a practical standpoint, we evaluate the real-time-factor (RTF) of the models on an Intel(R) Xeon(R) E5-2643 v4 CPU. The runtime is implemented based on LibTorch so as to conveniently deploy Pytorch models for production scenarios. Only one thread is used for CPU threading and TorchScript inference.
\section{Results and analysis}
\label{sec:typestyle}
\begin{table*}[htp]
	\caption{Results of ASV Conformer on VoxCeleb \& CN-Celeb. LSA means Length-Scaled Attention. For the configure, 6L-256D-4H-2Sub means a Conformer with 6 layers (L) blocks, 256 attention dim (D), 4 attention heads (H), and 2-factor subsampling (2Sub, default:4Sub)}
	\vskip 2pt
	\label{table:base}

   \resizebox{\linewidth}{!}{
	\begin{tabular}{llccccccc}

	\toprule
    \multirow{2}{*}{Model}  & \multirow{2}{*}{Configure}  &\multirow{2}{*}{Prarams} & \multirow{2}{*}{RTF}& VoxCeleb-O & VoxCeleb-E &  VoxCeleb-H & \multicolumn{2}{c}{CNC-Eval} \\
    & & & & EER(\%) & EER(\%) & EER(\%) & EER(\%) & $\text{minDCF}_{0.01}$ \\
    \midrule
    ECAPA-TDNN & C1024 & 16.0M &0.071 &0.925 & 1.231 & 2.321 & 9.45 & 0.5059 \\
    \quad + asnorm\cite{Matejka2017AnalysisOS}  &   &  & &0.856 & & & & \\ 
    original paper\cite{2020ECAPA}  &   &  & &0.87 & & & & \\ 
    \hline
    Conformer & 6L-256D-4H (w/o LSA) & 18.8M &0.025 &1.143 & 1.321 & 2.359 & 8.71 & 0.4747 \\
              & 6L-256D-4H           &       &     &1.026 & 1.280 & 2.278 & 8.39 & 0.4748 \\
              & 6L-256D-4H-2Sub & 22.5M &0.070 &\textbf{0.915} & \textbf{1.177} & \textbf{2.034} & \textbf{8.30} & \textbf{0.4504} \\
              & 12L-256D-4H  & 34.2M &0.030 &1.133 & 1.382 & 2.463 & 9.40 & 0.5195 \\
              & 6L-512D-8H  & 46.4M &0.080 &1.356 & 1.492 & 2.571 & 8.38 & 0.4826 \\
    \bottomrule
	\end{tabular}
	}
	\vspace{-1em}
\end{table*}
\subsection{Results of Conformer on VoxCeleb and CN-Celeb}
In this section, we generalize Conformer to ASV system. Table~\ref{table:base} presents the performance of different networks on VoxCeleb and CN-Celeb. We reproduced the 1024 channels ECAPA-TDNN as baseline system. Score Normalization \cite{Matejka2017AnalysisOS} (asnorm) here is just for a fair comparison with original paper and the results suggest the confidence of our baseline. For better compatibility with general ASR encoder setting, the attention dimension and head have two types (i.e., 256D-4H and 512D-8H). The efficacy of LSA is shown in the first two rows. LSA automatically adapts attention weight distribution, and generalizes model to various length inputs. Despite more parameters, the deeper 12 layers and wider 512D-8H networks fail to get further improvement, probably because 6L-256D-4H has enough model capacity to model the size of VoxCeleb and CN-Celeb. More complex structures are accompanied by harder convergence or overfitting. Compared with 4-factor, 2-factor subsampling generates longer/richer feature map before Conformer blocks, significantly improves performance, with the trade of computation cost.

From the aspect of inference speed, due to the quadratic-complexity of self-attention mechanism, 2-factor subsampling increases the RTF by 2$\times$-3$\times$ than 4-factor. The 512D also leads to about 4 times more computation compared to 256D. At last, compared with ECAPA-TDNN, 6L-256D-4H-Sub with comparable model size and RTF shows ideal improvement in EERs. While 6L-256D-4H achieves the remarkable relative reduction in RTF of 64.8\%, with desirable performance sacrifice.

\subsection{Effect of SAM training}
\label{sec:sam}
When it comes to an overparameterized circumstance, the 6L-512D-8H-6Sub\footnote{6L-512D-8H-6Sub matches the open source pretrained ASR model.} in Table~\ref{tab:sam} obtains unsatisfactory performance on VoxCeleb-O. A large regularization of 0.2 alleviates model overfitting, and a comparable efficacy can be achieved through SAM training. What's more, SAM training combined with appropriate regularization further improves the model generalization. 

To investigate the effective of ASR transferring, we downloaded the available GigaSpeech ASR checkpoint \cite{gigas} as a pretrained encoder. During training, we observed that the model transferred from the ASR encoder yields a lower value of training loss than training from scratch. However, the EER on test set increases from 1.431\% to 1.521\% as shown in the fifth and last line of Table~\ref{tab:sam}. It seems that prior ASR classification increases the risk of 
falling into a sharp minima on the travel from ASR to ASV, leading to suboptimal quality. SAM aims to seek out parameters whose entire neighborhoods have both low loss value, preventing model from dramatic shaking by inputs perturbation. Thus, we attempted to introduce SAM training to provide robustness. As a result, SAM apparently improved model generalization as the EER value decreases from 1.521\% to 1.218\%. Compared with standard ASV Conformer, ASR tranferring achieved 11.22\% relative reduction in EER under the same training procedure. Note that we didn't 
pre-train a 256D-4H ASR model on GigaSpeech, and ASR transferring is further discussed in sec~\ref{sec:transfe}.
\begin{table}[htp] 
\vspace{-1.5em}
	\caption{Effect of SAM training. ASR transfer means transferred from an ASR encoder trained on GigaSpeech.}
	\vskip 2pt
	\label{tab:sam}
    \centering
	\begin{tabular}{lccc}
	\toprule
	\multirow{2}{*}{Model} &\multirow{2}{*}{SAM} &\multirow{2}{*}{weight decay} & VoxCeleb-O \\
	    &  &   & EER(\%) \\
	\midrule                           
	     6L-256D-4H       &  \XSolidBrush &  0.05  & 1.026    \\ 
	                      & \Checkmark  &  0.05     & \textbf{0.983}    \\ \hline
	     6L-512D-8H-6Sub   & \XSolidBrush  &  0.05     & 1.771    \\
	      & \Checkmark  &  0.05     & 1.459    \\
	     & \XSolidBrush  &  0.2     & 1.431    \\
	     & \Checkmark  &  0.2     & 1.372    \\	 
	     \quad + ASR transfer  & \Checkmark   &  0.2 & \textbf{1.218}   \\
	       & \XSolidBrush  &  0.2     & 1.521    \\	                    
        \bottomrule	     
	\end{tabular}
\vspace{-2em}
\end{table}
\subsection{Effect of ASR transferring}
\label{sec:transfe}
Experiments conducted on CN-Celeb further reports the effect of ASR transferring. All transferred models are applied SAM. The 6L-256D-4H is chosen as baseline system. As shown in Table~\ref{trans}, Conformers of all different configurations benefit greatly from the ASR pretrained encoder, proving that our proposed method is effective enough for integrating ASR information. As WenetSpeech is much larger than Multi-CN, a more robust ASR encoder contributes better performance. When applying full ASR encoder to ASV system, the EER is improved by -1.582\% but the minDCF is degraded by +0.0617. This result might be due to the rear layers of the ASR encoder being too biaed towards ASR classification, making it more difficult for the network to learn speaker representation. Compared with training from scratch, transferring 6 blocks from the WenetSpeech ASR encoder achieves 11.56\% relative reduction in EER and 6.76\% relative reduction in minDCF respectively on CNC-Eval, revealing the capabilities of Conformer to model ASR and ASV tasks.
\begin{table}[htp]
\vspace{-2.0em}

	\caption{Performance of ASR transferring on CN-Celeb.}
	\vskip 2pt
	\label{trans}
    \centering
	\begin{tabular}{lccc}
	\toprule
	\multirow{2}{*}{Configure} &\multirow{2}{*}{pretrain ASR} & \multicolumn{2}{c}{CNC-Eval} \\
	                           &                         &  EER(\%) & $\text{minDCF}_{0.01}$ \\
	\midrule                           
	     6L-256D-4H            &  ---                    &  8.39     & 0.4748    \\ 
	                           & Multi-CN                &  7.95     & 0.4534    \\
	                           & WenetSpeech             &  \textbf{7.42}     & \textbf{0.4427}    \\ \hline
	     12L-256D-4H           &   ---                   &  9.40     & 0.5195    \\
	                           & WenetSpeech                &  7.82     & 0.5812    \\ \hline
	     6L-512D-8H               &   ---                   &  8.38     & 0.4826    \\
	                           & WenetSpeech             &  7.83     & 0.4551    \\
        \bottomrule	     
	\end{tabular}
\vspace{-2em}
\end{table}



\section{Conclusions}
\label{sec:coclu}
To bridge the gap between ASR and ASV, we concentrate on applying a unified architecture Conformer in ASV system. First of all, we adopt LSA and SAM training on ASV Conformer system to improve the model generalization ability. LSA enables model to generate to variable lengths inputs and SAM seeks a flatter loss landscape, proved to be effective especially when the model is severely overfitted. Experiments conducted on Voxceleb \& CN-Celeb indicated that the Conformer is well-suited for ASV task, and achieved competitive performance compared with the popular ECAPA-TDNN. Moreover, a simply ASR transferring method is introduced to integrate ASR information. ASR transferring outperformed standard ASV Conformer, gave a relative improvement of about 11\% in EER on both VoxCeleb and CN-Celeb. Unfortunately, the underlying relationship between ASR and ASV remains unclear and should be further investigated in future. In addition, we provide a runtime to better deploy models for production. The RTF of ASV Conformer evaluated on runtime environment verified its industrial value. Last but not least, this work leaves plenty room to extend advanced exploration of attention-based technology to ASV system, which will benefit ASV tasks both in industry and academia.

\section{Acknowledgements}
This work was funded by the National Natural Science Foundation of China (Grant No.61876160 and No.62001405) and in part by the Science and Technology Key Project of Fujian Province, China (Grant No. 2020HZ020005).

\vfill\pagebreak
\bibliographystyle{IEEEtran}
\bibliography{strings,refs}

\end{document}